\documentclass[
aps,epsfig,nofootinbib,
superscriptaddress,showpacs,floatfix]{revtex4}
\usepackage{graphicx}

\newcommand{\beq}{\begin{eqnarray}}
\newcommand{\eeq}{\end{eqnarray}}
\newcommand{\moms}{{\rm MOM}}
\newcommand{\moma}{\widetilde{\rm MOM}}
\newcommand{\msbar}{\overline{\rm MS}}
\newcommand{\LMSb}{\Lambda_{\msbar}}
\newcommand{\Lmoma}{\Lambda_{\moma}}
\newcommand{\Lmoms}{\Lambda_{\moms}}

\def\eq#1{Eq.~(\ref{#1})}

\begin{document}
\setcounter{page}{1}

\title{Three-gluon running coupling from lattice QCD at $N_f=2+1+1$: a consistency check of the OPE approach.}
\author{Ph.~Boucaud} 
\affiliation{Laboratoire Physique Th\'eorique, 
Universit\'e de Paris XI; B\^atiment 210, 91405 Orsay Cedex; France}
\author{M.~Brinet}
\affiliation{Laboratoire de Physique Subatomique et de Cosmologie, CNRS/IN2P3/UJF; 
53, avenue des Martyrs, 38026 Grenoble, France}
\author{F.~De Soto}
\affiliation{Dpto. Sistemas F\'isicos, Qu\'imicos y Naturales, 
Univ. Pablo de Olavide, 41013 Sevilla, Spain}
\author{V.~Morenas}
\affiliation{Laboratoire de Physique Corpusculaire, Universit\'e Blaise Pascal, CNRS/IN2P3 
63177 Aubi\`ere Cedex, France}
\author{O.~P\`ene}
\affiliation{Laboratoire Physique Th\'eorique, 
Universit\'e de Paris XI; B\^atiment 210, 91405 Orsay Cedex; France}
\author{K.~Petrov}
\affiliation{Laboratoire de
l'Acc\'el\'erateur Lin\'eaire, Centre Scientifique d'Orsay; B\^atiment 200, 91898 ORSAY Cedex, France}
\author{J.~Rodr\'{\i}guez-Quintero}
\affiliation{Dpto. F\'isica Aplicada, Fac. Ciencias Experimentales; 
Universidad de Huelva, 21071 Huelva; Spain.}

\begin{abstract}
We present a  lattice calculation of the renormalized running coupling constant in symmetric
(MOM) and asymmetric ($\widetilde{\rm MOM}$) momentum substraction schemes including $u$, $d$,
$s$ and $c$ quarks in the sea. An Operator Product Expansion dominated by the dimension-two
$\langle A^2\rangle$ condensate is used to fit the running of the coupling. 
We argue that the agreement in the predicted $\langle A^2\rangle$ condensate for both schemes is a strong
support for the validity of the OPE approach and the effect of this non-gauge invariant condensate
over the running of the strong coupling.

\begin{flushright}
LPT-Orsay 13-75\\
UHU-FT/13-10 \\
\end{flushright}
%
%

\end{abstract}


\maketitle

\section{Introduction}

The running of QCD coupling constant is one of the key ingredient for the confrontation of the 
experimental results to the perturbative expressions. The value of $\alpha_{\rm QCD}$ at
a given scale or alternatively the parameter $\Lambda_{\rm QCD}$, which controls the perturbative
running, can be extracted from experimental results. The running coupling constant can also be
computed from lattice QCD calculations by a large variety of methods (see, for instance, the 
{\it Particle Data Group} review~\cite{Beringer:1900zz}). Among these methods, we will pay 
attention to those based on the lattice determination of QCD Green 
funtions~\cite{Alles:1996ka,Boucaud:1998bq}. All of them need for a coupling to be 
nonperturbatively defined in a MOM-type scheme by fixing the 
QCD propagators (two-point Green functions) and one particular three-point Green-function 
for a chosen kinematical configuration to take, after renormalization, their tree-level result at 
the renormalization scale.

The analysis of the running for a so-defined coupling at intermediate energies, roughly from $3$ to $10$ GeV, 
deserves great interest as it provides with a privileged room for the confrontation of lattice nonperturbative 
results to perturbation theory, at any order, where the nature and impact of nonperturbative corrections 
can be studied (see ref.~\cite{Boucaud:2011ug} for a recent review). For instance, the study of the 
coupling defined from the  asymmetric ($\moma$) three-gluon vertex and estimated from quenched 
lattice data revealed the main role of nonperturbative power corrections to account for its 
running~\cite{Boucaud:2000ey}. Then, in a vast series of 
papers~\cite{Boucaud:2000nd,Boucaud:2001st,DeSoto:2001qx,Boucaud:2005rm,Boucaud:2005xn,
Boucaud:2008gn,Blossier:2010ky,Blossier:2010vt,Blossier:2011tf,Blossier:2012ef,Blossier:2013te}, 
some of us have exhaustively proven that Wilson's Operator Product Expansion (OPE) provides 
a general framework to include non-perturbative contributions, and that its application to QCD couplings 
for several renormalization schemes allows a coherent and simple explanation of the running 
obtained from the lattice for momenta as low as $\sim 2-3  {\rm GeV}$. The leading OPE contribution 
has been shown to result from the non-vanishing condensate of the gauge-dependent dimension-two 
local operator $A^2$~\cite{Lavelle:1992yh}, which, in the last decade, received profuse attention within  
the context of the so-called {\it refined Gribov-Zwanziger} approach~\cite{Dudal:2007cw,Dudal:2008sp,Dudal:2010tf} 
but also in many others (see~\cite{Gubarev:2000nz,Kondo:2001nq,Verschelde:2001ia,
Dudal:2002pq,Boucaud:2002nc,RuizArriola:2004en,Megias:2005ve,RuizArriola:2006gq,Megias:2009mp,Vercauteren:2011ze}).

Among the MOM schemes that have been studied, that defined for the
ghost-gluon vertex with zero incoming ghost momentum (called T-scheme)
has been extensively exploited in the last few years, due mainly to a well-known Taylor's result~\cite{Taylor:1971ff} 
whereby the proper ghost-gluon vertex renormalization constant for this scheme is proven to be exactly 
one in Landau gauge. Thus, the MOM T-scheme coupling  can be computed only from 
ghost and gluon propagators, without involving a three-point function. 
The latter allows for a very precise determination of $\alpha_s$ in a range of momenta
which makes possible to get an accurate estimate of $\Lambda_{\rm QCD}$ that, for realistic 
unquenched lattice simulations, succesfully compares to 
its value from experiments~\cite{Blossier:2011tf,Blossier:2012ef,Blossier:2013nw}. 

In any other scheme, the renormalized coupling requires
the lattice evaluation of a vertex function, i.e., a three point
correlation function. This is the case for the coupling defined from the three-gluon 
vertex~\footnote{The three-gluon vertex has been also the object of a recent study~\cite{Binosi:2013rba} grounding a 
QCD effective charge definition within the framework of the background field method and the pinching technique~\cite{Aguilar:2006gr,Binosi:2009qm}.}, where one can cook out as many different renormalization
schemes as there are possible kinematical configurations. As the signal for a three-point correlation 
function, suffering from stronger statistical fluctuations, is much harder to 
be extracted from lattice simulations than the one for two-point functions, 
the precision so attained is not comparable with the one achieved when using 
the T-scheme coupling. 
The interest of computing $\alpha_s$ in different schemes is therefore not
to obtain a precise value of $\Lambda_{\rm QCD}$ but, rather, to test the OPE framework 
and to gain thus some insight into the nature of the nonperturbative corrections.

In this paper we present the lattice evaluation of the MOM QCD coupling defined through the 
three gluon vertex for two different kinematical configurations: the symmetric (three equal momenta) and 
asymmetric (one vanishing momentum) ones. The high-statistics ensemble (800 configurations) of lattice 
gauge fields we exploit takes into account the dynamical generation of up, down, strange and charm 
quarks ($N_f=2+1+1$). This leaves us with two main "{\it aces}" for our game:
(i) the Wilson coefficients for the leading contribution in the OPE of the two couplings, as will be seen, 
differ very much from each other; and (ii) the perturbative running is very reliably known as the 
same $N_f=2+1+1$ lattice configurations provides, {\it via} the T-scheme coupling determination, 
with an accurate estimate of $\Lambda_{\rm QCD}$~\cite{Blossier:2013nw}, compatible with PDG world 
average~\cite{Beringer:1900zz}, that can be used here. We put ourselves in a near unbeatable position 
to check the OPE framework, as the nonperturbative contributions supplementing the perturbative running 
to account for the lattice data of both couplings can be properly isolated and compare to each other.  
One can see then if they differ as much as OPE predicts.

The structure of the paper is as follows: in section \ref{sec:mom} the renormalization schemes
and lattice setup used are described. In section \ref{sec:ope} a reminder of the OPE results 
has been included. Finally in section \ref{sec:res} our main results are presented and we concluded 
in section \ref{sec:conc}.

\section{Renormalization schemes.}
\label{sec:mom}

The starting point for this calculation shall be the gauge configurations produced
by the European Twisted Mass collaboration (ETMC) for $2+1+1$ dynamical quark flavors 
that provide a realistic description of the QCD dynamics including heavy flavours. 
These gauge field configurations, after fixing Landau gauge,
allow to compute the renormalized running coupling in momentum substraction schemes.
In particular we will focus on the coupling defined from three-gluon vertices.

\begin{figure}[h!]
\begin{center}
\includegraphics[width=0.2\textwidth]{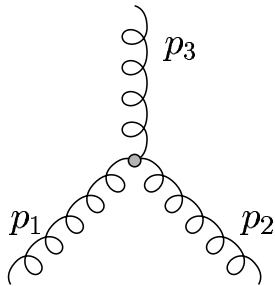} 
\caption{Three gluon vertex. MOM renormalization prescription implies that full three point function behaves as a renormalized
coupling (grey circle) times the three outgoing renormalized propagators.}
\label{fig:vertex}
\end{center}
\end{figure}

The three-gluon vertex (Fig.~\ref{fig:vertex}) can be computed from the lattice 
for any momenta $p_1$, $p_2$ and $p_3$ satisfying $p_1+p_2+p_3=0$. In particular, 
we will concentrate on the symmetric three gluon vertex ($p_1^2=p_2^2=p_3^2$) and the
asymmetric one ($p_3=0$ and therefore $p_2=-p_1$). The renormalized coupling can be straightrowardly defined
from gluon propagators and vertices (a detailed description of the procedure can be found in \cite{Boucaud:1998bq}). The renormalized coupling is defined by:
\begin{equation}
g_R(\mu^2) = \frac{Z_3^{3/2}(\mu^2) G^{(3)}(p_1^2,p_2^2,p_3^2)}{\left(G^{(2)}(p_1^2)G^{(2)}(p_2^2)G^{(2)}(p_3^2)\right)^{1/2}} 
\end{equation}
where $\mu^2$ is the renormalization scale, to be fixed for each renormalization scheme, $G^{2}(p^2)$ is the bare gluon propagator extracted from the lattice:
\begin{equation}
G^{(2)}(p^2) = \frac{\delta_{ab} g^{\mu\nu}}{3(N_C^2-1)} \langle \tilde{A}_\mu^a(p)  \tilde{A}_\nu^b(-p)\rangle\ ,
\end{equation}
$Z_3(\mu^2)=\mu^2 G^{(2)}(\mu^2)$ is the gluon field renormalization constant, 
and $G^{(3)}(p_1^2,p_2^2,p_3^2)$ is the scalar function extracted from three gluon vertex
${G^{(3)}}^{a b c}_{\mu\nu\rho}(p_1,p_2,p_3)=\langle \tilde{A}_\mu^a(p_1)  \tilde{A}_\nu^b(p_2) \tilde{A}_\rho^c(p_3)\rangle$
This scalar function is defined as the coefficient of the tree level tensor and is obtained after projecting the vertex
onto the adequate tensor as described in \cite{Boucaud:1998bq}.

This procedure allows to compute the running coupling $\alpha(\mu^2)=\frac{g_R^2(\mu^2)}{4\pi}$ in momentum
substraction schemes both from the symmetric three gluon vertex (MOM) and from the asymmetric one ($\widetilde{\rm MOM}$). 

The symmetric vertex requires the three momenta $p_1$, $p_2$ and $p_3$ to satisfy the constrain $p_1+p_2+p_3=0$ 
simultaneously with $p_1^2=p_2^2=p_3^2$ which is rather rare in the lattice. It means that there are rather few momenta
where the vertex can be evaluated. For the asymmetric one, the constrain is less restrictive and the vertex can be evaluated
at any lattice momenta $p$.

As mentioned above, the two and three-point gluon Green functions will be computed from lattice  gauge field
configurations simulated at $N_f$=2+1+1 by the ETM collaboration \cite{Baron:2010bv,Baron:2011sf}. The details 
of the computation can be found in \cite{Blossier:2011tf} and references therein. We have exploited here a set of $800$ configurations for a lattice volume $48^3$x$96$ and $\beta=2.10$, where the lattice parameters 
$\kappa_c=01563570$, $\mu_l=0.002$, $\mu_\sigma=0.120$, $\mu_c=0.385$, 
are fixed so that light quark mass is set to $\sim 20 {\rm MeV}$ and the strange and 
charm are set to $\sim 95 {\rm MeV}$ and $1.5 {\rm GeV}$ respectively (see \cite{Blossier:2011tf} and references therfein for the details of the simulations). According to Ref.~\cite{Blossier:2013nw}, the lattice spacing corresponding to this set-up is  
$a(2.10)=0.0583(11) {\rm fm}$.

Contrarily to the continuum ones, the lattice scalar functions do not depend only on the momentum squared $p^2$, due to 
the lattice discretization. Indeed, the lattice discretization breaks the $O(4)$-symmetry introducing an spurious dependence on
the invariants of the group $H(4)$. These $O(4)$-breaking lattice artefacts can be efficiently removed
by using the so-called $H(4)$-extrapolation procedure~\cite{Becirevic:1999uc,Becirevic:1999hj,deSoto:2007ht}. This method works efficiently for gluon propagator
and asymmetric vertex, where there is a high number of momenta at which the Green function can be evaluated. 
The correction of lattice
artifacts for the symmetric vertex is not so efficient due to the lower number of lattice momenta and the fact that it depends
on three momenta instead of one. This introduces a limitation for the larger momenta that can be used in the schema
${\rm MOM}$ that, in practice, implies a smaller fitting window than in $\widetilde{\rm MOM}$ scheme. As will be seen in next section, we take momenta below $a(2.10)p \simeq 1.6$ (around 5.5 GeV, in physical units) for the fit in the $\moma$ case, while the fitting window is restricted only to momenta below $a(2.10)p \simeq 1.3$ (around 4.5 GeV) for the fit in MOM, to avoid the noise induced by the non-properly-cured lattice artefacts.

\section{OPE nonperturbative predictions.}
\label{sec:ope}

The running of the strong coupling constant with momentum, obtained from QCD perturbation theory corrected by a nonperturbative leading OPE power contribution, can rather generally read~\cite{Blossier:2010ky,Blossier:2011tf}
\beq\label{eq:NPalpha}
\alpha_R(\mu^2)
\ = \ 
\alpha^{\rm pert}_R(\mu^2)
\ 
\left( 
 1 + \frac{c_R}{\mu^2} \
\left( \frac{\alpha^{\rm pert}_R(\mu^2)}{\alpha^{\rm pert}_R(q_0^2)}
\right)^{1-\gamma_0^{A^2}/\beta_0} 
R\left(\alpha_R^{\rm pert}(\mu^2),\alpha_R^{\rm pert}(q_0^2) \right) 
\frac{g^2_R(q_0^2) \langle A^2 \rangle_{R,q_0^2}} {4 (N_C^2-1)}
+ o\left(\frac 1 {\mu^2}\right) \rule[0cm]{0cm}{0.75cm}
 \right) \, , 
\eeq
where the subindex R specifies any particular renormalization scheme and $\alpha^{\rm pert}$ gives the running 
behaviour perturbatively obtained from the integration of the QCD beta function at that R scheme, 
\beq\label{eq:beta}
\frac{d}{d\ln{\mu^2}} h_R \ = \ - \left( \beta_0 h_R^2 
+ \beta_1 h_R^3 + \beta_2^R h_R^4 + \dots \right)
\eeq
with $h_R=\alpha_R(\mu^2)/(4\pi)$ and $\beta_0=11-2/3 N_f$, $\beta_1=102-38/3 N_f$, being scheme-independent coefficients.  
The result for $\alpha^{\rm pert}$ from the integration of \eq{eq:beta} and its conventional perturbative inversion, in terms of 
momenta and the QCD scale $\Lambda_R$, can be found in \cite{Beringer:1900zz}. 
Within the bracket, $c_R$ is given by the tree-level Wilson coefficient contribution, 
$\gamma_0^{A^2}$ is the first coefficient for the local operator $A^2$ anomalous dimension, determining the 
Wilson-coefficient leading-logarithm contribution, 
\beq
1-\gamma_0^{A^2}/\beta_0 = \frac {27}{132 - 8 N_f} 
\ ,
\eeq
which is found to be scheme-independent; and $R(\alpha,\alpha_0)$ encodes the higher-order logarithmic corrections for the leading Wilson coefficient~\cite{Blossier:2013te}. In refs.~\cite{Blossier:2010ky,Blossier:2012ef,Blossier:2013nw}, \eq{eq:NPalpha} particularized to the MOM T-scheme accounted very accurately for the running of the corresponding lattice data with momenta. 
This allowed for a precise determination of $\Lambda_T$, and hence $\LMSb$ pretty in agreement with the PDG~\cite{Beringer:1900zz} "{\it world average}".

Hereupon, we will mainly concentrate on the running coupling renormalized in both MOM 
and $\widetilde{\rm MOM}$ schemes, for which the three-loop beta coefficients are known~\footnote{The four-loop 
beta coefficient appears also to be known in the $\widetilde{\rm MOM}$ case}. 
For $N_f=4$, one is left with
\beq\label{eq:beta2}
\beta^{\moma}_2 = \ 814.56 \ , \ \ \ \ 
\beta^{\moms}_2 = \ 641.16 \ ;
\eeq
and with the following ratios 
\beq\label{eq:rats}
\frac{\LMSb}{\Lmoma} = \ 0.443 \ , \ \ \ \ 
\frac{\LMSb}{\Lmoms} = \ 0.463 \ ,
\eeq
that can be exactly obtained from the first non-trivial coefficient for the expansion of $\moms$ and $\moma$ couplings in 
terms of the $\msbar$ one. 
The tree-level Wilson coefficients for the strong coupling in both schemes have been also studied~\cite{Boucaud:2001st,DeSoto:2001qx} and their computation gives
\beq\label{eq:tree}
c_{\moma} = 3 \ , \ \ \ \ 
c_{\moms} = 9 \ .
\eeq
It is worthwhile to recall that, in the $\widetilde{\rm MOM}$ case, as a consequence of the soft gluon 
field in the three-gluon Green function defining the vertex, \eq{eq:NPalpha} only results after 
the factorization of a leading higher-dimension condensate~\footnote{The anomalous dimension first coefficient for the lower dimension operator in the $\widetilde{\rm MOM}$-case OPE expansion have been proven 
to differ from that of $A^2$ only in a negligible way~\cite{DeSoto:2001qx}.} in the OPE, 
induced by a vacuum insertion approximation~\cite{DeSoto:2001qx} (the same vacuum insertion approximation 
have been proven to work for the OPE expansion of the ghost-ghost-gluon Green function~\cite{Boucaud:2011eh}).
The function $R(\alpha,\alpha_0)$ is related to the $A^2$ anomalous dimension but also depends on the 
scheme we used to define the coupling~\cite{Blossier:2013te}. For the MOM T-scheme~\cite{Boucaud:2008gn}, as the coupling 
can be directly related to gluon and ghost propagators involving no three-point Green function, 
it has been computed at the $O(\alpha^4)$-order~\cite{Blossier:2010ky,Blossier:2013te}. Its computation is nevertheless 
cumbersome when dealing with three-gluon Green function is needed. Thus, in the following, we will 
take $R(\alpha,\alpha_0)=1$ and will work at the leading-logarithm approximation.

\section{Results.}
\label{sec:res}

Then, \eq{eq:NPalpha} can be fitted to the lattice data, with 
$g^2\langle A^2 \rangle$ as a free parameter, for both $\moms$ and $\moma$ couplings, 
with their perturbative predictions obtained by the integration of the beta function 
with the coefficients given in \eq{eq:beta2} and the ratios of $\Lambda$'s in \eq{eq:rats}. 
We will take $\Lambda_{\msbar}=314$ MeV, as an input from 
ref.~\cite{Blossier:2013nw}\footnote{It should be noted that \cite{Blossier:2013nw} applied now a very recent 
result for the lattice scale setting~\cite{Silvano:2013nw}, which slightly differs from that applied in 
refs.~\cite{Blossier:2011tf,Blossier:2012ef}, shifting down all the dimensionful quantities.} 
where, as above mentioned, the MOM T-scheme coupling is computed from the 
lattice and confronted with \eq{eq:NPalpha}, properly particularized (ref.~\cite{Blossier:2013nw} upgrades 
the previous results of refs.~\cite{Blossier:2011tf,Blossier:2012ef}). 

\begin{figure}
\begin{center}
\begin{tabular}{cc}
\includegraphics[width=0.5\textwidth]{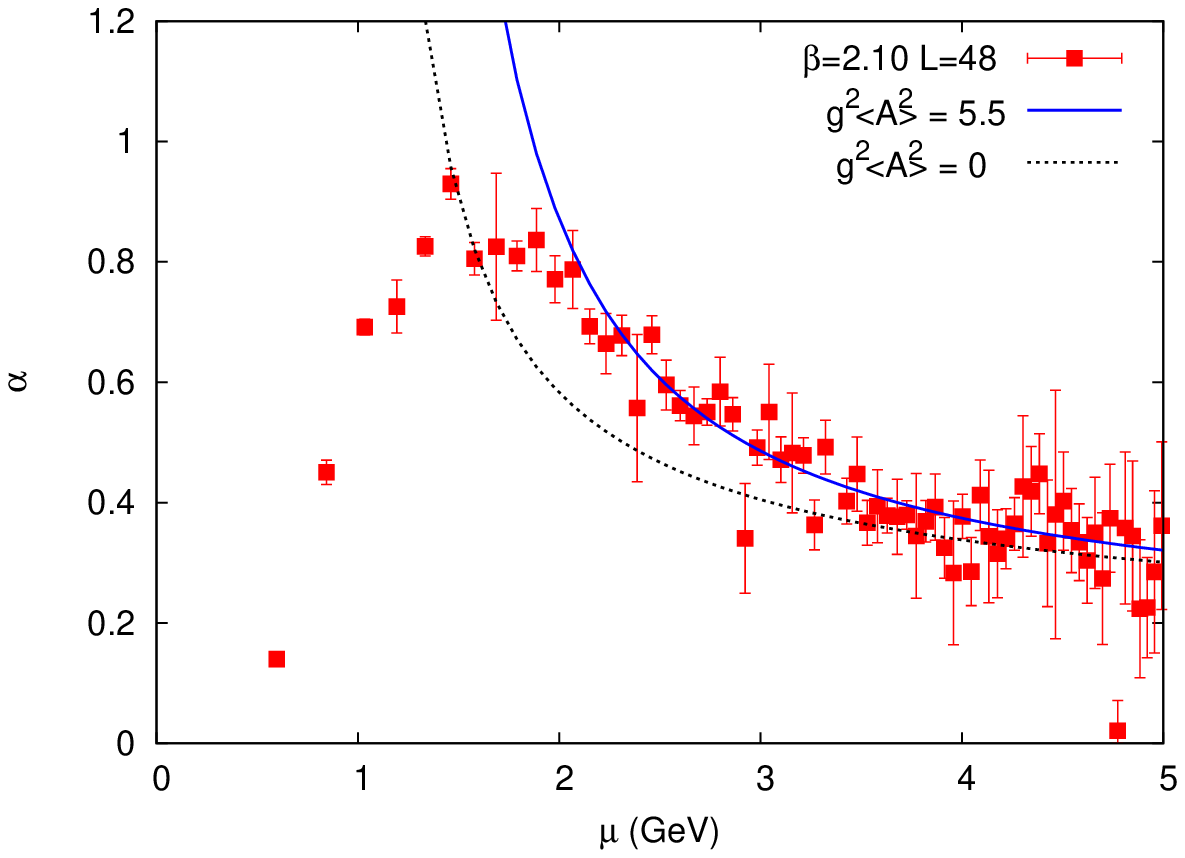} 
&
\includegraphics[width=0.5\textwidth]{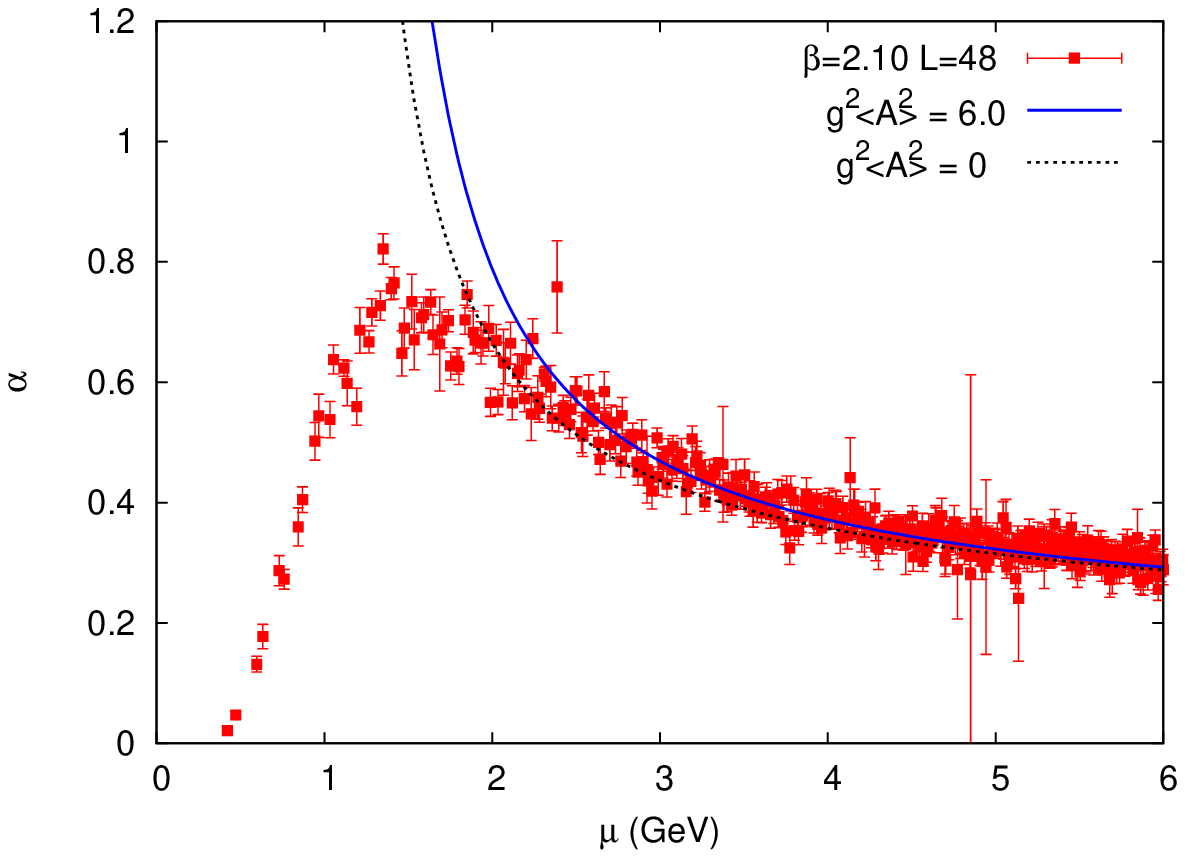} 
\end{tabular}
\caption{(Color online) Lattice data for $\alpha_{\rm MOM}(\mu)$ [left] and $\alpha_{\widetilde{\rm MOM}}(\mu)$ [right] vs the momenta $\mu$ in ${\rm GeV}$. The full line shows, in both cases, the non-perturbative fit discussed in the text while the dashed line stands for the corresponding perturbative running.}
\label{fig:mom}
\end{center}
\end{figure}

Figs.~\ref{fig:mom} shows the lattice results for the running coupling constant 
and the best fits with \eq{eq:NPalpha} and $g^2 \langle A^2\rangle$ from Tab.~\ref{tab:fits}, 
in the $\moms$ and $\moma$ schemes. 
Within the framework of OPE and SVZ sum-rules approach, 
the gluon condensate needs to take similar values in the OPE for the two couplings. The 
agreement in the values of the extracted condensates is therefore a strong indication of the validity of this approach, at least for a window of momenta not lying in the deep IR domain. 
As the nature of the OPE condensates is the object of a recent controversy~\cite{Brodsky:2010xf,Chang:2011mu}, It is worth to point out that our check is validating the 
sum-rules factorization but does not tell necessarily anything about the nature of the condensates. 

\begin{table}[h]
\begin{tabular}{|| c | c | c ||c||}
\hline
\hline
 & $\rm MOM$ & $\widetilde{\rm MOM}$ & T-scheme \\
\hline
$g^2\langle A^2\rangle\ (\rm GeV^2)$ & 5.5 $\pm$ 0.8 & 6.0 $\pm$ 1.3 & 5.5 $\pm$ 1.0 \\
\hline
$\chi^2/d.o.f$ & 0.76 & 0.89 & \\
\hline
\hline
\end{tabular}
\caption{Condensate $g^2\langle A^2\rangle_{R,\mu_0}$ renormalized at $\mu_0=10 {\rm GeV}$ extracted from the best fit of the lattice running couplings 
$\alpha_{\rm MOM}(\mu)$ and  $\alpha_{\widetilde{\rm MOM}}(\mu)$. For comparison, the MOM T-scheme result, estimated from ref.~\cite{Blossier:2013nw} data as explained in the text, is also included.}
\label{tab:fits}
\end{table}

The values of the condensates obtained in this paper are sizably larger than the ones reported in
\cite{Blossier:2011tf,Blossier:2012ef,Blossier:2013nw} for the Taylor scheme. It is however important to note 
that, in the Taylor scheme analysis of those papers, the beta function is expanded at the four-loop 
level for its integration and 
that the Wilson coefficient has been computed up to the $O(\alpha^4)$-order. Here, in our analysis of 
${\rm MOM}$ and  $\widetilde{\rm MOM}$ couplings, we have only used a three-loop beta function and Wilson 
coefficient at the leading order. Indeed, the effect of including higher orders in either the perturbative part of 
\eq{eq:NPalpha} or the Wilson coefficient is well known to reduce the value of the 
condensate~\cite{Boucaud:2008gn,Pene:2011kg}.   
Alternatively, for the sake of a consistent comparison, we repeated the analysis of Ref.~\cite{Blossier:2013nw} 
under the same approximation level (the best which can be here coherently attained) 
applied for the current one: the perturbative coupling expanded only up to three-loops and the Wilson coefficient 
kept at the leading-logarithm approximation. One then obtains 
\beq\label{eq:g2A2}
g^2\langle A^2 \rangle=5.5(1.0) \ \mbox{\rm GeV}^2 \ ,
\eeq
always at the renormalization point $\mu=10$ GeV. This last result 
happens to be exactly the same as the one for $\moms$ and to lie in the same ballpark as that for $\moma$, as 
can be seen in Tab.~\ref{tab:fits}. Indeed, one can also apply 
both $\Lambda_{\overline{\rm MS}}$ from \cite{Blossier:2013nw} and $g^2\langle A^2\rangle$ from \eq{eq:g2A2} 
to \eq{eq:NPalpha} and, without free parameters to be fitted, account for the lattice data for $\moms$ 
and $\moma$ couplings with $\chi^2/$d.o.f. that would be then 1.37 for the 
former and 0.76 for the latter. 

It is worthwhile to emphasize that refs.~\cite{Blossier:2011tf,Blossier:2012ef,Blossier:2013nw} exploited 
the same (800) lattice configurations for the gauge fields at a bare coupling, $\beta=2.1$, here analysed, 
but also configurations at $\beta=1.90$ for three different light-quark twisted masses (500 each) 
and at $\beta=1.95$ (150). The latter gives us the grounded conviction that lattice artefacts are properly 
under control in obtaining $\Lambda_{\overline{\rm MS}}$ and $g^2\langle A^2 \rangle$ from the Taylor coupling, 
as done in \eq{eq:g2A2}. Therefore, that \eq{eq:NPalpha} successfully describes the $\moms$ and $\moma$ coupling 
data for a given momentum window, with the same $\Lambda_{\overline{\rm MS}}$ and $g^2\langle A^2 \rangle$, 
strongly indicates that lattice artefacts appear to be negligible also for them, after $H(4)$-extrapolation, 
within such a window.

\begin{figure}
\begin{center}
\begin{tabular}{cc}
\includegraphics[width=0.47\textwidth]{acomp.eps} &
\includegraphics[width=0.5\textwidth]{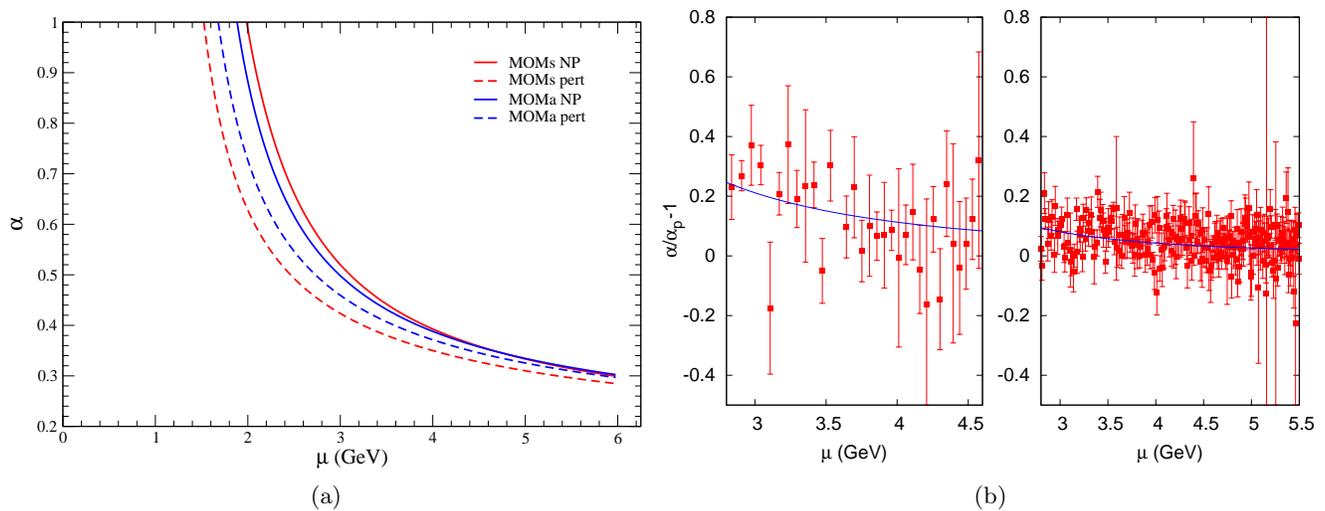} \\
(a) & (b)
\end{tabular}
\caption{[Color online] (a) For the sake of comparison, the perturbative (dashed) and OPE nonperturbative $\moms$ (red) and $\moma$ (blue) predictions are displayed together. (b) Ratio $\alpha/\alpha^{Pert}-1$ for the ${\rm MOM}$ (left) and  $\widetilde{\rm MOM}$  (right) 
schemes vs the momenta $\mu$ in ${\rm GeV}$. The lines show the non-perturbative fit according to Eq.~(\ref{eq:fit}) discussed in the text.}
\label{fig:ratios}
\end{center}
\end{figure}

That the condensates obtained from both $\moms$ and $\moma$ takes the same value is a demanding 
result, as the Wilson coefficient is three times larger for the former than for the latter. This 
implies that deviations from the perturbative behaviour should be very different in both cases, being 
consistent with the ratio of 3 given by \eq{eq:tree}. This can be seen in Fig.~\ref{fig:ratios}.a and, 
otherwise presented, as follows:

\eq{eq:NPalpha}, in the leading logarithm approximation, and \eq{eq:tree} left us with:
\beq\label{eq:check}
\frac{\frac{\displaystyle \alpha_{\moms}(\mu^2)}{\displaystyle \alpha^{\rm pert}_{\moms}(\mu^2)} - 1}
{\frac{\displaystyle \alpha_{\moma}(\mu^2)}{\displaystyle \alpha^{\rm pert}_{\moma}(\mu^2)} - 1} \ = \ 
\frac{c_{\moms}}{c_{\moma}} + O\left(\alpha,\frac{1}{\mu^2}\right) \ = \ 
3 + O\left(\alpha,\frac{1}{\mu^2}\right)\ ,
\eeq
which provides with a very demanding consistency check for the OPE and SVZ sum-rules approach, which is 
totally equivalent to the compatibility of condensates in Tab.~\ref{tab:fits}.   
To perform this check, we need to compute \eq{eq:check}'s l.h.s. from 
the lattice data for the $\moma$ and $\moms$ three-gluon coupling and their perturbative predictions 
obtained again by the integration of the beta function with the coefficients given in \eq{eq:beta2}, 
the ratios of $\Lambda$'s in \eq{eq:rats} and  $\Lambda_{\msbar}=314$ MeV from 
ref.~\cite{Blossier:2013nw}. However, as the lattice momenta for $\moma$ and 
$\moms$ differ, and aiming at employing as large a statistics as possible, \eq{eq:check}'s l.h.s. will be 
indirectly computed by fitting both numerator and denominator, within as large as possible a momentum 
domain for each, to
\beq\label{eq:fit}
\frac{\alpha_R(\mu^2)}{\alpha_R^{\rm pert}(\mu^2)} - 1 \ = \ a_R \ 
\frac{\left(\alpha_R^{\rm pert}(\mu^2)\right)^{0.27}}{\mu^2} \ ,
\eeq
as suggested by \eq{eq:NPalpha}, with $a_R$ as the only free parameter to be fitted. This can be seen 
in the plots of Fig.~\ref{fig:ratios}.b, where it clearly appears that, as the running given by 
\eq{eq:fit}'s r.h.s. is near the same for both schemes, $a_{\moms}$ is to be rather larger than 
$a_{\moma}$. Thus, once both parameters are fitted, they can be applied to compute \eq{eq:check}'s 
l.h.s.,
\beq\label{eq:check2}
\frac{\frac{\displaystyle \alpha_{\moms}(\mu^2)}{\displaystyle \alpha^{\rm pert}_{\moms}(\mu^2)} - 1}
{\frac{\displaystyle \alpha_{\moma}(\mu^2)}{\displaystyle \alpha^{\rm pert}_{\moma}(\mu^2)} - 1} \simeq 
\frac{a_{\moms}}{a_{\moma}} = \frac{2.28(34) \ \mbox{\rm GeV}^2}{0.83(19)\ \mbox{\rm GeV}^2} = 2.7(8) \ ,
\eeq
that compares remarkably well with \eq{eq:check}'s r.h.s. evaluated through the OPE results given by \eq{eq:tree}. 
It should be noticed that, in \eq{eq:check2}, $\alpha_{\moms}/\alpha_{\moma}=1+O(\alpha)$ has been applied, as 
corresponds to our approximation level (see \eq{eq:check}).
We performed the fit in a momentum window $p\in (2.8, 4.5) {\rm GeV}$ in the $\moms$ case and 
$p\in (2.8, 5.5) {\rm GeV}$ in $\moma$. The errors for the fitted parameters, $a_R$, have been computed by applying 
the jackknife procedure, and propagated then into the final result for the ratio. 

\section{Summary and conclusions}
\label{sec:conc}

We used lattice gauge field configurations, generated with four twisted-mass dynamical quark flavours (two light degenerate and two heavy) within the framework of ETM collaboration, to compute the running of the QCD coupling constant, 
$\alpha_s$, defined from the symmetric and asymmetric three-gluon vertices.
This leads us to two different MOM-type renormalization schemes for the coupling, where their running with momenta, roughly from 
3 to 6 GeV, has been described only after supplementing the well-known perturbative prediction with a 
non-perturbative correction dominated by a non-vanishing dimension-two gluon condensate in Landau gauge, 
$\langle A^2\rangle$. 

As can be seen in Fig.~\ref{fig:ratios}.a, the perturbative estimate for the coupling in $\moma$ scheme is larger than the one in $\moms$. This is a direct consequence from the way their three-loop $\beta$ coefficients in \eq{eq:beta2}  
and the ratios in \eq{eq:rats}  compare to each other. Contrarily, 
the OPE nonperturbative leading correction for $\moms$ is predicted to be three times larger than 
for $\moma$, irrespectively with the value of $\langle A^2 \rangle$. Being so that altogether, 
for a sufficiently large condensate, the relative strength between the two nonperturbative couplings 
will result reversed with respect to the perturbative case. We indeed found the lattice data for the $\moma$ and $\moms$ coupling  to follow this reversed pattern. We also measured the ratio between their nonperturbative corrections and found it to be 2.7(8), strongly supporting the OPE approach to account for the nonperturbative contributions. It is worth to recall that the prediction for this ratio is only relying on the Shifman-Vainshtein-Zakharov  technology~\cite{Shifman:1978bx,Shifman:1978by} to compute the OPE Wilson coefficients and the universality of the involved condensate. 

Finally, the value of the condensate is also found to be consistent with that resulting from the analysis of the running 
of the coupling in the MOM T-scheme, in ref.~\cite{Blossier:2013nw}, after restricting the calculation there  
to the same order than here.  This provides thus with additional support for the 
very accurate estimate of $\Lambda_{\msbar}$ obtained therein.

\bibliography{refs}

\end{document}